\documentclass[aip,letterpaper,twocolumn,reprint,floatfix,nofootinbib]{revtex4-2}
\usepackage{amsmath}
\usepackage{graphicx}
\usepackage[usenames,dvipsnames]{color}
\usepackage{booktabs}
\usepackage{color}

\begin{document}

\title{On measuring the bending modulus of lipid bilayers with cholesterol}
\author{John F. Nagle}
\email[E-mail: ]{nagle@cmu.edu}
\affiliation{$^1$Department of Physics, Carnegie Mellon University, Pittsburgh, PA 15213, USA}

\date{\today}


\begin{abstract}
Regarding the effect on the bending modulus of adding cholesterol to lipid bilayers, recent results using neutron spin echo and nuclear magnetic resonance relaxation methods that involve linear transport properties have conflicted with earlier results from purely equilibrium experiments that do not involve linear transport properties. A general discussion indicates how one can be misled by data obtained by methods that involve linear transport properties.  It is then shown specifically how the recent neutron spin echo results can be interpreted to agree with the earlier purely equilibrium experimental results, thereby resolving that conflict.  Regarding the nuclear magnetic resonance relaxation method, it is noted that current interpretation of the data is unclear regarding the identity of the modulus that is involved, and an alternative interpretation is explored that does not disagree with the results of the equilibrium experiments.
\end{abstract}



\maketitle


\section{Introduction}
A decade ago, four independent groups reported the unanticipated and still surprising result that the bending modulus $K_C$ of DOPC does not increase with addition of cholesterol.  Bassereau’s group\cite{sorre2009} and Baumgart’s group\cite{tian2009bending} studied tubules, Dimova’s group\cite{gracia2010} analyzed shape fluctuations (SA) in giant unilamellar vesicles as well as vesicle deformation in electric fields, and my group\cite{pan2008chol,pan2009chol} analyzed X-ray diffuse scattering (XDS) from stacks of bilayers.  My group also systematically studied the effect of cholesterol on other standard lipids with the expected result that $K_C$ increased dramatically for bilayers of DMPC lipid that has saturated hydrocarbon chains.  For mono-unsaturated SOPC, $K_C$ increased, but less.  And $K_C$ did not increase for C22:1PC, which like DOPC, is di-unsaturated. 

 In striking contrast, a recent paper reports that the bending modulus of DOPC increases threefold with increasing cholesterol concentration up to 50\%.\cite{chakraborty2020}  The two experimental techniques that pertained to the bending modulus were neutron spin echo (NSE) and nuclear magnetic resonance relaxation (NMRR).  This development has made this into a controversial topic.\cite{nagle2021needless,ashkar2021reply} This paper will attempt to reconcile this controversy.

The four methods in the first paragraph are quite different from each other. But they share the feature that the interpretations of their data do not involve non-equilibrium quantities and therefore rely only on the firm principles of equilibrium statistical mechanics. In contrast, NSE and NMRR data basically report decay rates which necessarily involve non-equilibrium quantities.  It is usually assumed that the lipid bilayer system can be treated in the linear transport regime of non-equilibrium statistical mechanics, and then viscosity and friction are quantities that have to be included for interpreting the data in addition to the static elastic moduli. While this potentially allows NSE and NMRR to obtain more information than the purely equilibrium methods, it also means that there are more quantities that have to be determined from limited data.  Most importantly, it means that the non-equilibrium statistical mechanical theory for interpreting the data has to be correct.  Given that non-equilibrium theory is more complex and less certain than equilibrium theory, it is clearly a warning that something may be wrong when those conclusions do not agree with results from four purely equilibrium methods.

The NSE and NMRR data are appealing because they clearly show that cholesterol slows down measured rates substantially.  However, this alone says nothing directly about the bending modulus. Section \ref{sec:Prop-2} reminds biophysicists of this. Section \ref{sec:NSE-3} reviews details of the NSE interpretation.  This interpretation has gradually evolved and it is accepted in this paper up to the final equation.  That equation had been modified some time ago for cholesterol in a non-NSE context.\cite{pan2009chol}  That modification fully reconciles the NSE data with the equilibrium results.  In contrast, the NMRR data are not so easily understood as is discussed in Section \ref{sec:NMRR-4}. The NMRR data interpretation is forty years old.  Arguments are given that, like the NSE interpretation, it too needs to evolve, and a suggestion is discussed for what that might look like.  In Section \ref{sec:MD-5} there is a brief discussion of molecular dynamics (MD) simulations.  A simulation that was performed to support the NSE and NMRR experimental results\cite{chakraborty2020} disagrees with an earlier simulation that agrees with the results of the equilibrium methods.\cite{Harries2018}

\section{Propaedeutics}
\label{sec:Prop-2}
The bending modulus is a static equilibrium quantity. That may sound strange to biophysicists because it is often measured from fluctuations, as in the XDS and SA methods, until it is appreciated which aspects of fluctuations yield purely equilibrium quantities and which require non-equilibrium quantities.  Of course, any system in equilibrium in a thermal environment at temperature T must, over time, repeatedly sample all its configurations $\xi$ with Boltzmann relative probabilities $exp(-E(\xi)/kBT)$ where $E(\xi)$ is the energy of state $\xi$. A system in thermal equilibrium with its environment at non-zero temperature cannot just sit in its lowest energy state; it must have equilibrium fluctuations.  That is consistent with the system being in its lowest free energy state.  In the case of the bending modulus of membranes, $K_C$ is determined by the SA and XDS methods from the average mean square equal-time equilibrium fluctuations in the Fourier coefficients of the height-height correlation functions.  The tubule method\cite{sorre2009,tian2009bending} and the electro-deformation method\cite{gracia2010} are actually more direct in that they measure bending under steady applied force rather than relying on fluctuations.

\begin{figure}
\includegraphics[width=1.0\columnwidth,angle=0]{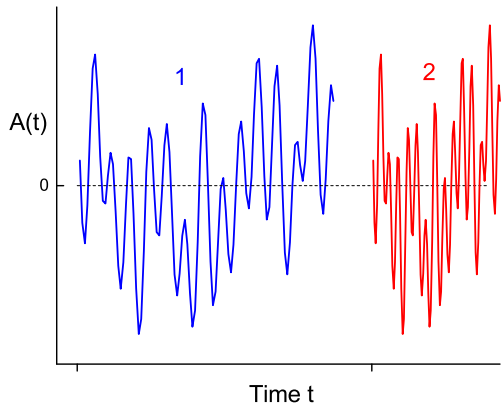}{}
	\caption[]{Schematic of the fluctuations A(t) in the dynamic amplitude  for system 1 (left) and system 2 (right).}
	\label{fig:PREg}
\end{figure}

It is important that biophysicists appreciate how and why the equilibrium fluctuations used to measure the bending modulus are distinct from dynamics.  Fig.\ref{fig:PREg} shows two dynamical traces corresponding to two systems, labelled 1 and 2, for the fluctuations in any general amplitude $A$ as a function of time.  The two traces are identical except that the slow trace occurs at half the speed of the fast trace.  Therefore, the time average of the mean square fluctuations ${\langle}A(t)^2{\rangle}_{t}$ is identical. Those are the fluctuations whose Fourier transforms are directly related to the bending modulus by the standard, well-known formula $K_C = k_BT/{\langle}A(q)^2{\rangle}q^4$ for bilayers under zero tension,\cite{boal2012book} so the bending modulus of the two systems must also be equal despite having unequal dynamics. 

Another example of this fundamental distinction is if the trace amplitudes illustrated in Fig.\ref{fig:PREg} were to report time-independent fluctuations in the lengths of two springs.  Then the spring constants, which are also equilibrium quantities like moduli, must also be equal. A likely difference in molecular nano scale systems, which can make spring 1 dynamically slower than spring 2, is viscous friction.  Viscosity is a non-equilibrium quantity in the linear transport category.  A system with higher viscosity that is pulled from its mean value will relax back more slowly than one that has the same energy landscape but with lower viscosity.  When the system spontaneously undergoes a thermal fluctuation from its average value, it likewise takes longer for the system to fluctuate back towards its average value so the dynamics are stretched out as in Fig.\ref{fig:PREg}. 

Of course, it should be remembered that the fundamental fluctuation-dissipation theorem\cite{kubo1985} importantly relates fluctuations to non-equilibrium transport quantities, but those are not the time-independent equal-time fluctuations ${\langle}(A(t))^2{\rangle}_{t}$ used for the equilibrium moduli.  Instead, they are the time dependent correlations ${\langle}A(t+\tau)A(t){\rangle}_t$ at different times $\tau$. The faster dynamics in system 2, namely the faster decay of ${\langle}A(t+\tau)A(t){\rangle}_t$ with $\tau$ would be related by the fluctuation-dissipation theorem to a non-equilibrium transport quantity such as a smaller viscosity in system 2 than in system 1.

This is a good place to comment on misconceptions about time scales of various experiments.  It was written that the time scale of the XDS experiment is long,\cite{chakraborty2020} but the actual XDS time scale is of the order of femtoseconds, far shorter than the nanosecond time scale of NSE, and therefore \textit{a priori} superior for obtaining the relevant $\tau$ = 0 correlations for the equilibrium bending modulus.  As in any experimental method data are accumulated over many events; XDS data are accumulated over a long time to obtain good statistics from many photons, each with its femtosecond time scale.  It should also be emphasized that there is no relevant time scale for the tubule method or the electro-deformation method, both of which are performed in the steady state. 

It has also been suggested  that NSE and NMRR are at the appropriate molecular length scale,\cite{chakraborty2020} but there is no separate bending modulus for shorter length scales. The classical Helfrich bending modulus is defined as the leading order of energy that is dominant at the longest length scale. At shorter length scales, consideration of other types of energy can become important.  It has become clear that molecular tilt plays a role at shorter length scales and that requires a separate tilt modulus.\cite{kozlov2000,kopelevich2007,watson2011thermal,doktorova2017}  Disentangling these two moduli experimentally is non-trivial, although it has been accomplished by XDS.\cite{jablin2014,nagle2017experimentally}  

Let us also introduce another example that might help illuminate the concepts.  Suppose that one increases the mass of a system in such a way that its energy landscape does not change.  Again, the relaxation time is longer and the thermal time-dependent fluctuations are slower because the random hits from the thermal environment result in smaller changes in velocity because of greater inertia of the system, but the average mean square equal-time fluctuations will not change over time and any equilibrium modulus will have the same value. This suggests a macroscopic example that may be intuitively useful as it can be perceived at the personal level.  Suppose you approach an unlatched door that is in the closed position.  If the door has a large mass, it will appear “stiffer” to you when you push to go through it than a less massive door. This perception of stiffness has nothing to do with the energy landscape of the doors; the corresponding moduli can be made equal by the usual (though somewhat complicated in practice) door closure hardware.  Instead, perceived door stiffness is related to non-equilibrium perturbations to the system, in this case your pushing the door.  Likewise, the more massive door will relax to closure more slowly with equal closure hardware.  Of course, the more rapidly you open the door, the more work you do, and that extra work is dissipated irreversibly.

These examples should alert biophysicists to an important distinction between physical quantities of interest that can be obtained from a system in equilibrium. Some quantities are static equilibrium moduli that can be obtained straightforwardly from time-independent quantities like ${\langle}A(t)^2{\rangle}_t$.  The static moduli suffice for describing cellular subsystems that effectively operate near equilibrium.  Other non-equilibrium quantities like viscosity and friction are related, rather less straightforwardly and limited to the linear regime, to measurement of the $\tau$ time-dependent ${\langle}A(t+\tau)A(t){\rangle}_t$ auto-correlation function.  When a cell process deviates from reversibility, non-equilibrium dissipative quantities become important, and the relative importance of the two types of quantities depends upon how far the process deviates from reversibility. Because different biological processes span a large range of time and length scales, the details of the actual biological process matter.  Assigning a single “effective” bending modulus from experiments that are unrelated to a specific biological process\cite{chakraborty2020} is likely to be misleading and impede physical understanding of biological processes.

\section{Reconciliation of NSE with equilibrium DOPC/cholesterol results}
\label{sec:NSE-3}

The NSE story is quite interesting.  In its experimentally available time window NSE provides the $q$ dependent relaxation rates for bending fluctuations $\Gamma(q)$ which rather non-trivial non-equilibrium theory interprets as         
\begin{eqnarray}
\Gamma(q)=0.025\left(\frac{k_BT}{K_C}\right)^{1/2}\left(\frac{k_B T}{\eta}\right)q^3,
\label{Eq:1-nse}
\end{eqnarray}
where $\eta$ is the solvent viscosity.\cite{ZG1996}  The presence of the non-equilibrium transport property $\eta$ in this interpretation is expected and reiterates the fact that NSE measures dynamics in a time window that does not permit direct measurement of ${\langle}A(t)^2{\rangle}_t$ from which the static bending modulus could be directly obtained.  While it was historically assumed that $K_C$ in Eq.(\ref{Eq:1-nse}) was the static equilibrium bending modulus, when the water value of $\eta$ was used, the extracted value of $K_C$ was much greater than any other measurement of $K_C$ and for a time a work-around used a value of $\eta$ roughly three times greater than the viscosity of water.\cite{yi2009bending,takeda1999nse} 

That problem was resolved for NSE analysis when Watson et al. developed a non-trivial quantitative theory on top of the previous theory to take into account internal friction in the membrane.\cite{watson2011intermediate}  The NSE theory then evolved to
\begin{eqnarray}
\Gamma(q)=0.025\left(\frac{k_BT}{K_{CD}}\right)^{1/2}\left(\frac{k_B T}{\eta}\right)q^3,
\label{Eq:2-nse}
\end{eqnarray}
This replaced $K_C$ in Eq.(\ref{Eq:1-nse}) with a dynamical bending modulus in the NSE time regime $K_{CD}$ given by\cite{seifert1994} 
\begin{eqnarray}
K_{CD}=K_C+h^2K_A,
\label{Eq:3-Brown}
\end{eqnarray}
where $K_A$ is the area compressibility modulus of the bilayer and $h$ is the distance from the bilayer midplane to the neutral surface of either monolayer of a symmetric bilayer.  The neutral surface, closely related to the pivotal plane, is an important conceptual property in membrane mechanics used by theorists.\cite{seifert1994,watson2011intermediate,campelo2014helfrich,hossein2020spontaneous}, It is the location in each monolayer where stretching is decoupled from bending.\cite{safran1994}  It has not been experimentally measured in bilayers, but is widely assumed to be located near the interface between the hydrocarbon chain region and the headgroups.  

It is interestingly deceptive that all the quantities on the right hand side of Eq.(\ref{Eq:3-Brown}) are equilibrium quantities.  Even though the derivation\cite{watson2011intermediate} involves non-equilibrium internal viscosity $\eta_s$ and a friction coefficient $b$, in the time and length scale regimes of NSE (but not in those of dynamic light scattering), these non-equilibrium transport quantities drop out of Eq.(\ref{Eq:3-Brown}), rather like the Cheshire cat leaving behind only an equilibrium smile.

It is illuminating to consider the relative magnitudes of the terms in Eq.(\ref{Eq:3-Brown}).  For typical values of $h$ and $K_A$,\cite{rawicz2000} the $h^2K_A$ term in Eq.(\ref{Eq:3-Brown}) is an order of magnitude greater than $K_C$.  This is consistent with the previously noted result that $K_C$ was an order of magnitude too large when the viscosity of water was used in Eq.(\ref{Eq:1-nse}).  Given this weak dependence of $K_C$ on the NSE measured $K_{CD}$, Eq.(\ref{Eq:3-Brown}) suggests that a more appropriate and important way to use NSE data would be to provide the first experimental measurement of the neutral surface $h$ in bilayers.  Values of $K_A$ only have 10\% uncertainties.\cite{rawicz2000}  There are larger uncertainties of 30\% in $K_C$ between different techniques,\cite{nagle2017experimentally} but that matters even less because $K_C$ itself makes such a small contribution to $K_{CD}$ in Eq.(\ref{Eq:3-Brown}).  (It should also be noted that, in contrast to the magnitude of $K_C$, the uncertainties in the cholesterol dependence of $K_C$ between different methods are much less than 30\%.)  If the theory\cite{watson2011intermediate} giving Eq.(\ref{Eq:2-nse}) and Eq.(\ref{Eq:3-Brown}) is accurate, then NSE could provide experimental values for the fundamental $h$ quantity. 

Instead, NSE practitioners have been prone to interpret their data as yet another way to obtain $K_C$.  To do this, a popular relation between $K_C$ and $K_A$ has been invoked\cite{nagao2017}
\begin{eqnarray}
K_A=\beta\frac{K_C}{(2D_c)^2} ,
\label{eq:4-KA}
\end{eqnarray}
where $2D_C$ is the hydrocarbon thickness of the bilayer.  The number $\beta$ has been variously given as 12 for coupled monolayers, 48 for uncoupled monolayers and 24 for the polymer brush model.\cite{rawicz2000}  Combining Eq.(\ref{eq:4-KA}) with Eq.(\ref{Eq:3-Brown}) then yields the formula 
\begin{eqnarray}
K_C = \frac{K_{CD}}{1+\beta(\frac{h}{2D_C})^2} ,
\label{eq:5-KCD}
\end{eqnarray}
This equation has been used to obtain $K_C$ by assuming values of $\beta$, $h$ and using experimental values for $2D_C$.  Although it was stated that the polymer brush model was used,\cite{nagao2017} the actual values used were $\beta$ = 48 and $h$ =  $D_C$.  Using these values in Eq.(\ref{eq:5-KCD}) and using the result to eliminate $K_{CD}$ in favor of $K_C$, one obtains an equation that has exactly the same form as Eq.(\ref{Eq:1-nse}) except that the numerical prefactor 0.025 is replaced by 0.0069.\cite{chakraborty2020,nagao2017} The point of recounting this history is to emphasize that the recent experimental determinations of $K_C$ using NSE rely on (i) Zilman-Granek theory,\cite{ZG1996} (ii) Watson-Brown theory,\cite{watson2011intermediate} and most importantly for what follows, (iii) an assumed relation between $K_C$ and $K_A$ in Eq.(\ref{eq:4-KA}). 

We are now ready to return to the controversial case of cholesterol in DOPC.  The NSE results that have led to the controversy have used Eq.(\ref{eq:4-KA}) with essentially the same values of $\beta$ and the ratio $h/D_C$ as cholesterol was added. Instead, let us bypass Eq.(\ref{eq:4-KA}) and examine what Eq.(\ref{Eq:3-Brown}) can tell us directly from experiment and by assuming plausibly constant values of $h$.  Then, the cholesterol dependence of $K_{CD}$ mainly depends on $K_A$.  The experimental $K_A$ of DOPC upon adding cholesterol increases by a factor of 3 with 50\% cholesterol.\cite{rawicz2008} The threefold increase with cholesterol in the dynamical bending modulus $K_{CD}$ that is obtained by NSE\cite{chakraborty2020} is therefore entirely consistent with the threefold experimental increase in $K_A$ and requires no increase in the minor contribution from the static equilibrium bending modulus $K_{C}$. This essentially resolves the controversy regarding NSE data.

But this controversy is insightful because the different behavior of $K_{A}$ and $K_{C}$ with cholesterol concentration in DOPC implies that Eq.(\ref{eq:4-KA}) does not hold for cholesterol.  This was already appreciated some time ago.  Evan Evans informed the authors of the XDS study that the basis of the polymer brush model was not satisfied when a rigid molecule like cholesterol is added and he suggested an alternative model for high 50\% cholesterol.\cite{pan2009chol}  Evans’s cholesterol model postulated that the mechanical properties are dominated by a stiff region in each uncoupled monolayer that has a length $\delta$ = 9 \AA~ corresponding to the length of the rigid hydrocarbon rings of cholesterol. This theory replaced Eq.(\ref{eq:4-KA}) with 
\begin{eqnarray}
K_A=12\frac{K_C}{\delta^2}
\label{eq:6-KA}
\end{eqnarray}
Combining with Eq.(\ref{Eq:2-nse}) gives
\begin{eqnarray}
K_C = \frac{K_{CD}}{1+12(h/\delta)^2} ,
\label{eq:7-KC}
\end{eqnarray}
If we assume that $h$ is the same as the monolayer hydrocarbon thickness $D_C$ = 15.3 Å at 50\% cholesterol,\cite{pan2009chol} then the denominator on the right had side of Eq.(\ref{eq:7-KC}) is about 35 for 50\% cholesterol.  In contrast, the denominator of Eq.(\ref{eq:5-KCD}) is 13 for no cholesterol.  This would require $K_{CD}$ to increase essentially by a factor of 3 in order for $K_{C}$ to increase at all.  Instead, it was noted\cite{pan2009chol} that an increase in $K_{A}$ in Eq.(\ref{eq:6-KA}) would be consistent with no increase in $K_{C}$ if $\delta$ were somewhat larger than 9 \AA.  A purely phenomenological calculation of $\delta$ that satisfies Eq.(\ref{eq:6-KA}) obtained a cholesterol dependence that sensibly interpolated to pure DOPC and Eq.(\ref{eq:4-KA}).  And the same phenomenology appeared to apply to SOPC bilayers with cholesterol.\cite{pan2009chol} 

Although it would be good to have a more fundamental microscopic theory to gain further insight into the effect of cholesterol on the mechanical properties of lipid bilayers, this section shows that there is no conflict between the NSE results and the older experimental results that reported no increase in $K_{C}$ as cholesterol is added to DOPC bilayers.

\section{NMRR and the DOPC/cholesterol controversy}
\label{sec:NMRR-4}
Nuclear magnetic resonance relaxation reports\cite{chakraborty2020} relaxation time $\tau$ which is a dynamical quantity involving ${\langle}A(t+\tau)A(t){\rangle}_t$.  As such, the data are related to a non-equilibrium quantity as is indeed shown by the presence of viscosity $\eta$ in the formula\cite{chakraborty2020} used to interpret the measured quantity C  
\begin{eqnarray}
C = \frac{3k_BT\sqrt{\eta}}{5{\pi}S_S^2\sqrt{2K^3}}.
\label{Eq:8-NMRR}
\end{eqnarray}
Interestingly, this $\eta$ was described as a membrane viscosity in contrast to the solvent viscosity in the NSE formula in Eq.(\ref{Eq:1-nse}).  The presence of the molecular order parameter $S_S$ emphasizes the small length scale of the method. The $K$ modulus in Eq.(\ref{Eq:8-NMRR}) was declared to be essentially equal to the bending modulus $K_C$ divided by the membrane thickness, although it was noted that $K$ was a single elastic constant, consistent with there being other possibilities.  The original paper that introduced $K$ declared that it could be derived from the general liquid crystal literature, but the specific references given were for the nematic phase.\cite{brownM1982} Since the nematic phase has no layers to undulate with a related bending modulus $K_C$, this prompts one to ask about the meaning of $K$ in the context of the continuum description that actually defines the moduli of membranes.  It might also be mentioned that Eq.(\ref{Eq:8-NMRR}) is based on the hypothesis that the slow NMR dynamics are due to collective motions; that hypothesis has been challenged by fitting NMR data to a non-collective model.\cite{pastor2008}

Both nematic and smectic phases have molecular tilt degrees of freedom.  This has become increasingly recognized in the membrane literature and the corresponding modulus in the collective continuum description is the tilt modulus $K_t$ that accompanies the tilt vector $\bf{m}$.\cite{kozlov2000} The relevant part of the free energy in this description is 
\begin{align}
\begin{split}
H = \frac{1}{2}\int_A d^2r\sum_{n}^{}
((K_C\left(\nabla^2_{r}u_n+\nabla_r\cdot\bf{m_n}\right)^2+\\\label{Eq:9-continuum}K_t{\bf{m_n}^2}+B(u_{n+1}-u_n)^2),
\end{split}
\end{align}
where the average bilayer normal is along the $z$ axis with fluctuations $u_n$ from the average position of the $n^{th}$ bilayer. For multilamellar systems with more than one membrane $n$, $B$ is the bulk compression modulus.  No derivation of the connection between this defining description of the moduli and the $K$ that appears in Eq.(\ref{Eq:8-NMRR}) has been published. 

It has been recognized that $B$ plays a role in NMRR analysis,\cite{halle1997} although it has been argued that it does not affect the analysis in the MHz range of multilamellar smectic systems because the length scale is limited to that of membranes undulating freely from their neighbors.\cite{nevzorov1997dynamics} Ignoring concerns of being able to discern gradual crossover between these length scales, let us focus on tilt in Eq.(\ref{Eq:9-continuum}) which has not yet been considered in Eq.(\ref{Eq:8-NMRR}).\cite{chakraborty2020,brownM1982,nevzorov1997dynamics,nevzorov1998lipid} It is well recognized that the effect of tilt on the Fourier undulation spectrum increases relative to the effect of bending as the length scale decreases toward the smaller molecular length scale with a crossover length $(K_C/K_t)$ of order 1 $nm$.\cite{kopelevich2007}  Given the recent emphasis placed on the small length scale of NMRR,\cite{chakraborty2020} it seems appropriate to consider that $K$ may be more related to the tilt modulus $K_t$ than to the bending modulus $K_C$.  

Let us consider the possibility that a proper derivation might confirm Eq.(\ref{Eq:8-NMRR}) but with a different interpretation for $K$.  The previous interpretation is that $K$ equals $K_C$ divided by hydrocarbon thickness 2$D_C$, and it will now be considered that $K$ = $K_t$ times 2$D_C$.  Using typical values of $10^{-19}$ J for $K_C$, 2.5 nm for $2D_C$, and 0.05 N/m for $K_t$ would replace $K$ = $4x10^{-11}$ N  by $12.5x10^{-11}$ N in Eq.(\ref{Eq:8-NMRR}).  For the same measured values of $C$ and $S_S$ in Eq.(\ref{Eq:8-NMRR}), this increase in $K$ by a factor of 3.1 would increase the membrane viscosity $\eta$ by a factor of $3.1^3$ = 30.  The recent paper did not report a value for viscosity obtained from NMRR to compare to the value for DOPC of $1.5x10^{-8}$ Pa.s.m obtained from NSE.\cite{chakraborty2020}  However, an earlier NMRR study\cite{nevzorov1998lipid} reported a bulk viscosity for DMPC of $\eta$ = 0.0707 Pa.s, which converts, after multiplying by the hydrocarbon thickness, to a membrane viscosity of $1.8x10^{-10}$ Pa.s.m, two orders of magnitude smaller than the NSE value for DOPC.  This unlikely large difference would be alleviated by multiplying by the factor of 30 obtained above by replacing $K$ = $K_C/2D_C$ by $K_t2D_C$.  Of course, these numbers are rough and depend upon an NMRR formula that hasn’t yet been derived, but they support the tentative suggestion that the tilt modulus is something to be considered in the interpretation of NMRR data.  

What is needed for the interpretation of NMRR data is a proper derivation of a relation to the moduli that are fundamentally defined in the continuum description in Eq.(\ref{Eq:9-continuum}). Such a formula would likely include the tilt modulus, the bending modulus, the bulk modulus $B$ for multilamellar dispersions, and maybe even the new splay-tilt coupling term.\cite{terzi2019} It had been previously noted that the single-modulus $K$ in Eq.(\ref{Eq:8-NMRR}) might consist of more than one modulus, but the suggested formula (Eq. 4.9)\cite{nevzorov1997dynamics} assigned the same $q$ dependence to all moduli.  A proper connection to the continuum energy definition in Eq.(\ref{Eq:9-continuum}) would necessarily assign different powers of $q$ to the different moduli.\cite{kopelevich2007}  Of course, viscosity and/or friction would also appear; note that there are both membrane viscosity $\eta_m$ and a separate friction term $b$ for intermonolayer slippage.\cite{watson2011intermediate,bingham2015} It might also be noted that a major improvement in the analysis of NSE data occurred when it was realized that the time dependence of the relaxation required a stretched exponential.\cite{ZG1996}  This possibility does not appear to have been considered in the analysis of NMRR data. 

It would likely be a daunting task for NMRR analysis to obtain the numerical values of all the properties that affect the data.  This task would be easier if some values could be fixed to those obtained from other experimental methods.  Measurements that only focuses on viscosity, like a newly developed one,\cite{faizi2021viscosity} could provide that value.  Equilibrium methods that only focus on $K_C$ could provide that value.  For multilamellar systems, the $B$ value is obtained by XDS.\cite{pan2009chol}  At this time, only the XDS method obtains experimental values for $K_t$ and that is empirically challenging because the lower end of the x-ray length scale is a bit larger than the molecular length scale.\cite{jablin2014,nagle2017experimentally} The added value of NMRR could be that it might be a second method for measuring $K_t$ as well as tying together the other quantities. Whether or not that will transpire, it is clear that the present assertion that cholesterol increases $K_C$ in DOPC bilayers\cite{chakraborty2020} is not firmly based on the NMRR data, so it is quite insufficient to overturn the original equilibrium results for the effect of cholesterol on the $K_C$ of DOPC lipid bilayers.\cite{sorre2009,tian2009bending,gracia2010,pan2008chol,pan2009chol} According to the tentative suggestion presented here for re-interpreting $K$ in Eq.(\ref{Eq:8-NMRR}), the NMRR data would indicate a plausible threefold increase in the tilt modulus $K_t$ as cholesterol is added to DOPC.

\section{Molecular dynamics}
\label{sec:MD-5}
It would be satisfying if the remarkably good agreement of the molecular dynamics simulations\cite{chakraborty2020} with the dynamical NSE $K_{CD}$ is simply because both are dynamical methods.  Unfortunately, that is not the case.  Molecular dynamics is truly dynamical in the sense that an MD simulation obtains trajectories in time. By calculating relaxation and time dependent autocorrelation functions, MD can obtain dynamical non-equilibrium properties by looking at $\tau$-dependent ${\langle}A(t+\tau)A(t){\rangle}_t$ fluctuations. But MD can also obtain equilibrium properties and many molecular dynamics studies go to considerable effort to ensure that simulations run long enough to achieve and then sample the equilibrium fluctuations. Then, time averaged mean square equal time ($\tau$=0) fluctuations ${\langle}A(t)^2{\rangle}_{t}$  can be extracted to provide equilibrium thermodynamic quantities that do not involve time or dynamics.  Like Monte Carlo methods, this use of molecular dynamics is a powerful way to sample the equilibrium ensemble and provide static equilibrium quantities of the ${\langle}(A(t))^2{\rangle}_{t}$ type.  The reason for emphasizing this is to avoid semantic confusion when referring to molecular dynamics simulations that obtain the equilibrium bending modulus in a way that does not use dynamical ${\langle}A(t+\tau)A(t){\rangle}_t$ quantities.  

The atomistic MD simulations that report an increase in bending modulus of DOPC with increasing cholesterol similarly obtain non-dynamical time averaged mean square equal time fluctuations ${\langle}A(t)^2{\rangle}_{t}$.\cite{chakraborty2020}   Therefore, this simulation should be obtaining $K_C$ not $K_{CD}$, so it is mysterious why the simulated $K_C$ cholesterol dependence agrees so well with the NSE $K_{CD}$ results.  Another perhaps related mystery is that this MD simulation group has used similar analysis methods to study the area compressibility modulus $K_A$ of each monolayer in a bilayer.\cite{doktorova2019KA} The counter intuitive result was reported that each monolayer in a symmetric bilayer has the same value of $K_A$ as the bilayer itself instead of the intuitively obvious and rigorously proven result that $K_A$ for a bilayer is the sum of the $K_A$ of its monolayers.\cite{nagle2019KA}  The analysis of both the simulations for $K_A$ and for $K_C$ used a real space method rather than the original, more commonplace, Fourier spectral analysis.\cite{edholm2000,goetz1999,venable2015mechanical}  However, a different real-space simulation study reported a slight decrease in the bending modulus with increasing cholesterol in DOPC,\cite{Harries2018}9 rather than a large increase, so it may not be the real space method \textit{per se} that accounts for these mysteries. While there should be no controversy regarding experimental results for DOPC with cholesterol, the simulations\cite{chakraborty2020} do remain controversial.  

\section{Summary and Afterthoughts}
\label{sec:SUM-6}
NSE and NMRR methods provide information about important non-equilibrium properties that could enrich our understanding of bilayers.  However, the non-equilibrium statistical mechanics required to interpret NSE and NMRR data makes those methods rather more problematic for obtaining the equilibrium bending modulus.  Indeed, the qualitative disagreement with purely equilibrium results for the case of cholesterol added to DOPC lipid bilayers indicates that there are flaws in the recent analyses.  Nevertheless, a relatively small modification to the last part of the recent NSE data analysis completely resolves that inconsistency.  It is to be hoped that consistency for the NMRR method can also be achieved by updating the formula used for interpreting those data.

Although it is clear experimentally that the equilibrium bending modulus $K_C$ of DOPC bilayers does not increase with up to 50\% cholesterol, this result remains surprising, especially since the bilayer becomes thicker as cholesterol is added.\cite{pan2009chol}  There is no \textit{ab initio} theory that explains why $K_C$ does not increase, either for DOPC or for the longer chain di-unsaturated C22:1PC lipid, whereas $K_C$ does increase with added cholesterol in SOPC that has one saturated chain and quite dramatically in DMPC that has both chains saturated.\cite{pan2009chol}  While this is not well understood theoretically, it might be biomedically relevant because animals require cholesterol and membranes have to be flexible, so a way to satisfy both requirements is to have lipids with unsaturated chains rather than having short saturated chains.\cite{nagle2013Faraday}  Asserting that there is a universal effect of cholesterol on all lipid bilayers\cite{chakraborty2020,ashkar2021reply} imposes an artificial consistency that is likely to impoverish our understanding of membrane mechanics.   Instead, it is to be hoped that the additional perspectives that can be brought to bear by a variety of properly interpreted experimental techniques will further our understanding of the effect of cholesterol on the biophysical properties of membranes.

Acknowledgements:  I thank Markus Deserno, Rumiana Dimova, Michihiro Nagao, Elizabeth Kelley and Evan Evans for helpful comments.

\section*{References}
\bibliography{refs}

\end{document}